\documentclass{article}
\usepackage{amsmath}
\usepackage{amssymb}
\usepackage{placeins}
\usepackage[pdftex]{graphicx}
\pdfoutput=1
\usepackage{geometry}
\title{{Colossal topological Hall effect at the transition between isolated and lattice-phase interfacial skyrmions}}
\author{M. Raju$^{{1\dagger}\ddagger}$, A.P. Petrovi\'c$^{1}$, A. Yagil$^{2}$, K.S. Denisov$^{3}$, N.K. Duong$^{1}$, B. G\"obel$^{4}$,\\
 E. \c{S}a\c{s}{\i}o\u{g}lu$^{4}$, O.M. Auslaender$^{2\star}$, I. Mertig$^{4}$, I.V. Rozhansky$^{3}$, C. Panagopoulos$^{1\dagger}$}

\date{}
\begin{document}
\maketitle

$^{1}$Division of Physics and Applied Physics, School of Physical and Mathematical Sciences, Nanyang Technological University, 637371 Singapore\\
$^{2}$Department of Physics, Technion, Haifa 32000, Israel\\
$^{3}$Ioffe Institute, Politekhnicheskaya 26, 194021 St.Petersburg, Russia\\
$^{4}$Institut f\"ur Physik, Martin-Luther-Universit\"at Halle-Wittenberg, 06099 Halle (Saale), Germany\\
$^\ddagger$Present address: Institute for Quantum Matter and Department of Physics and Astronomy, Johns Hopkins University, Baltimore, MD, USA\\
$^\star$Present address: Neuroscience Institute and Tech4Health Institute, NYU Langone Health, New York, NY\\
 $^\dagger$Correspondence: mraju@ntu.edu.sg/christos@ntu.edu.sg

%\linenumbers
%\def\linenumberfont{}

\section{Abstract}
The topological Hall effect is used extensively to study chiral spin textures in various materials. However, the factors controlling its magnitude in technologically-relevant thin films remain uncertain. Using variable-temperature magnetotransport and real-space magnetic imaging in a series of Ir/Fe/Co/Pt heterostructures, here we report that the chiral spin ﬂuctuations at the phase boundary between isolated skyrmions and a disordered skyrmion lattice result in a power-law enhancement of the topological Hall resistivity by up to three orders of magnitude. Our work reveals the dominant role of skyrmion stability and configuration in determining the magnitude of the topological Hall effect.

%%%%%%%%%%%%%%%%%%%%%%%%%%%%
\section{Introduction}
Magnetic skyrmions are topologically-charged nanoscale spin textures which form due to competition between spin-rotating and spin-aligning magnetic interactions. In thin film heterostructures these magnetic interactions can be finely tuned via the multilayer geometry and composition, rendering skyrmion-hosting films ideally suited for spintronic applications~\cite{Bogdanov2001,fert2017magnetic,MoreauLuchaire2016,Woo2016,Soumyanarayanan2017}. One promising route towards functionalising skyrmions in metallic systems is to utilise their intrinsic magnetoelectric coupling, which is manifested by a topological Hall effect ($THE$)~\cite{Neubauer2009,Nagaosa2013,Spencer2018}. Charge carriers moving through a skyrmion spin texture experience an emergent magnetic field ($B_{eff}$) associated with the spin winding of a skyrmion. The transverse deflection of charge carriers interacting with $B_{eff}$ results in a $THE$ \cite{Nagaosa2013}.
Provided that each skyrmion remains stationary with respect to incident charge carriers, an array of skyrmions will exhibit a topological Hall resistivity $\rho^{THE}$, given by

\begin{equation}
\rho^{THE}=P\cdot R^{'}_{0}\cdot(n_{sk}\cdot \Phi_0),
\end{equation}
where $P$ is the spin polarization of the charge carriers, $R^{'}_{0}$ is the Hall coefficient representing the effective charge density contributing to the $THE$ (usually taken as the classical Hall coefficient $R_0$), $B_{eff}\equiv~n_{sk}\cdot \Phi_0$ is the emergent field associated with a given skyrmion density $n_{sk}$, and $\Phi_{0}=h/e$ is the magnetic flux quantum, with $h$ Planck's constant and $-e$ the electron charge. This phenomenon is distinct from the classical and anomalous Hall effects, which are proportional to the applied magnetic field $H$ and magnetization $M(H)$, respectively \cite{Nagaosa2013}. 
Experimentally, $\rho^{THE}(H)$ can be identified as the residual between the total measured Hall resistivity $\rho_{yx}(H)$ and a fit to the classical $R_0(H)$ and anomalous $R_{S}M(H)$ Hall resistivities namely $\rho_{yx}^{fit}(H)=R_0H+R_\mathrm{S}M(H)$~\cite{raju2019evolution} (where $R_\mathrm{0}$ and $R_\mathrm{S}$ are the classical and anomalous Hall coefficients respectively).
The anomalous Hall resistivity $\rho^{AHE}\equiv~R_{S}M(H)$ can be described by a superposition of terms with linear and quadratic dependences on the longitudinal resistivity $\rho_{xx}$, corresponding to the skew scattering and side jump terms respectively \cite{Spencer2018}. However, the variation of $\rho_{xx}(H)$ in these multilayers is extremely small ($<0.01\%$ for fields up to magnetic saturation at $H_S$ and $<0.17\%$ for fields up to $\pm5$~T \cite{Soumyanarayanan2017,raju2019evolution}). Consequently, the treatment of $\rho^{AHE}$ as $[a\cdot \rho_{xx}(H)+b\cdot \rho_{xx}^{2}(H)]M(H)$ or simply $R_{S}M(H)$ has negligible influence on the estimated $\rho^{THE}$ \cite{raju2019evolution}. A detailed scaling analysis of $\rho^{AHE}$ with $\rho_{xx}$ as a function of temperature $T$ ($5-300$~K) and $H$ for these multilayers can be found in Refs. \cite{Soumyanarayanan2017,raju2019evolution}, together with a discussion of the validity and reproducibility of the estimated $\rho^{THE}$.

Using equation ($1$) and our experimetally-determined $\rho^{THE}$, one can hence estimate $n_{sk}$ from an electrical transport measurement as:

\begin{equation}
n_{sk}(THE)=\mid\rho^{THE}\mid\div\mid(P\cdot R_0\cdot \Phi_0)\mid.
\end{equation}

$n_{sk}$ may also be measured directly using real-space imaging techniques such as magnetic force microscopy ($MFM$)~\cite{raju2019evolution,wang2018ferroelectrically,vistoli2019giant}, magnetic transmission X-ray microscopy ~\cite{Woo2016,Soumyanarayanan2017} or Lorentz transmission electron microscopy~\cite{he2018evolution}. Comparing these transport and imaging approaches can yield evidence for adiabatic transport if $n_{sk}(THE) \approx n_{sk}(MFM)$\cite{Neubauer2009,Nagaosa2013,Spencer2018}, non-adiabaticity if $n_{sk}(THE) < n_{sk}(MFM)$~ \cite{denisov2016electron,DenisovPRB2018}, or alternatively reveal enhanced transverse scattering mechanisms if $n_{sk}(THE) > n_{sk}(MFM)$ \cite{denisov2016electron,Ishizuka2018,kato2019colossal,ishizuka2019theory}. 
%%%%%%%%%%%%%%%%%%%%%%%%%%%%%%%%%%%
%%%%%%%%%%%%%%%%%%%%

In bulk non-centrosymmetric materials which exhibit stable skyrmion lattices, the values of $n_{sk}$ estimated from transport and imaging are in good agreement \cite{Neubauer2009,Nagaosa2013,Spencer2018}.  However, this is not the case for thin film multilayers with an interfacial Dzyaloshinskii-Moriya interaction ($DMI$), which can be tuned to exhibit isolated or dense skyrmion configurations. Large, conflicting values for $\rho^{THE}$ have been reported in magnetic thin films, corresponding to $n_{sk}(THE)$ orders of magnitude larger than $n_{sk}(MFM)$~\cite{Soumyanarayanan2017,zeissler2018discrete,wang2018ferroelectrically,vistoli2019giant,he2018evolution,Matsuno2016}.  

Determining the mechanism leading to this extraordinary disagreement is crucial for understanding the electrical response of chiral spin textures and their detection in devices. Recently, a valuable clue has been provided by predictions~\cite{Ishizuka2018,kato2019colossal,ishizuka2019theory} and observations~\cite{wang2019spin,Cheng2019} of spin fluctuation-induced effects on charge transport in non-coplanar magnets.  It is hence plausible that quantum or thermal fluctuations may influence the Hall response of materials with a chiral instability.
%%%%%%%%%%%%%
%%%%%%%%%%%%%%%%%%

In this work, using temperature-dependent Hall transport and $MFM$, we track the evolution of $THE$ across the transition between isolated skyrmions and a disordered skyrmion lattice. We find that isolated skyrmion configurations produce a larger $THE$ than dense arrays of skyrmions, with an enhancement of up to three orders of magnitude at the transition. Our data reveals a power-law behavior in $n_{sk}(THE)/n_{sk}(MFM)$ which we interpret in terms of chiral spin fluctuations. Following universal scaling laws, we extract the critical exponents governing this phase transition ~\cite{leonov2018crossover, leonov2016properties,siemens2016minimal}.
%%%%%%%%%%%%%%%%%%%%%%%
%%%%%%%%%%%%%%%%%%%%%%%%%%%%%

\section{Results}
Figure 1 shows measurements of $\rho^{THE}(H)$ at $300$~K for a set of Ir/Fe($x$)/Co($y$)/Pt multilayers. $MFM$ images acquired at $H$ maximizing $\rho^{THE}(H)$ display spin configurations ranging from isolated skyrmions (Fig.1a-c) to dense, disordered skyrmion lattices (Fig.1d-g) (see section 1 of the supplementary information for the evolution of magnetic textures). 
This evolution in skyrmion configuration is driven by the $T$-dependent stability parameter $\kappa\equiv~\pi D/4\sqrt{AK_{eff}}$, which describes the competition between the three key magnetic interactions: exchange coupling ($A$), magnetic anisotropy ($K_{eff}$) and $DMI$ ($D$). Our multilayer Ir/Fe($x$)/Co($y$)/Pt stacks allow us to systematically tune $\kappa$ (and hence $n_{sk}$) via the Fe/Co layer thickness ratio: $x/y$~$<0.5$ yields $\kappa<1$, whereas $\kappa\geq1$ for $x/y$~$\geq0.5$ \cite{Soumyanarayanan2017}. Varying $T$ provides an additional handle to tune $\kappa$ for a given Fe/Co ratio, $\kappa$ increases with $T$ due to $T$-dependent magnetic interactions (see section $2$ of the supplementary information). At $H=0$, these chiral magnetic films exhibit a labyrinthine stripe domain phase. Under a finite $H$, this transforms into a metastable skyrmion phase if $\kappa<1$ (Fig.1(a)-(c)), or a disordered skyrmion lattice phase if $\kappa\geq{1}$ (Fig.1(d)-(f)). For $\kappa\geq{1}$, the skyrmion lattice dissolves into isolated skyrmions before a uniformly polarized ferromagnetic ($FM$) phase develops for $H>H_S$. The isolated skyrmion phase emerges between a lattice and a $FM$ phase; its appearance can be regulated by $\kappa$, $T$, $H$, or a combination of these parameters (see section 1 of the supplementary information). Indeed, in chiral magnetic films the transformation of a polarized FM phase into an array of isolated skyrmions and ultimately a skyrmion lattice occurs via a nucleation-type second-order phase transition~\cite{leonov2018crossover,leonov2016properties,siemens2016minimal}. 
%%%%%%%%%%%%%%
%%%%%%%%%%%%%%%%
  
We find that isolated skyrmion configurations (Fig.1a-c) consistently generate a larger $\rho^{THE}(H)$, despite their small $n_{sk}(MFM)$. Contrary to expectations from equation (1), dense skyrmion arrays (Fig.1d-f) with larger $n_{sk}(MFM)$ typically display a smaller $\rho^{THE}(H)$. The generic large $\rho^{THE}(H)$ for isolated skyrmions can be further confirmed using the evolution of a dense skyrmion array with increasing $H$, as shown in Fig.1d-f. Here, $\rho^{THE}(H)$ rises as the skyrmion lattice dissolves into isolated skyrmions before reaching a $FM$ phase with $\rho^{THE}(H)\approx 0$. To further probe this discrepancy between transport and imaging experiments, we evaluate $\Delta_{\rho/M}\equiv~n_{sk}(THE)/n_{sk}(MFM)$  ~\cite{zeissler2018discrete,raju2019evolution,vistoli2019giant,he2018evolution,shao2019topological,Matsuno2016}, which quantifies the effective topological charge contributing to the measured $THE$, then compare this quantity for different skyrmion configurations imaged by $MFM$. As $n_{sk}(MFM)$ increases, $\Delta_{\rho/M}$ is systematically reduced (Fig.~1g). Hence  for a skyrmion lattice we have $n_{sk}(THE) \approx n_{sk}(MFM)$ and for isolated skyrmions $n_{sk}(THE)>n_{sk}(MFM)$. The strong influence of the skyrmion configuration on $\rho^{THE}$ underlines the crucial role of $\kappa$ in determining the magnitude of $\rho^{THE}(H)$. We therefore examine the transition between isolated and dense arrays of skyrmions, and the influence of $\kappa$ on the measured $THE$.
%%%%%%%%%%%%%
%%%%%%%%%%%%%
  
Figure 2a depicts the $T-\kappa$ space and hence the wide range of spin textures which we can experimentally access by varying $T$ ($5-300$~K), the Fe/Co thickness ratio and the number of repeats of [Fe/Co] in a multilayer stack. Using our $MFM$ images we identify the transition (blue shaded region in Fig.~2a) between isolated skyrmions and a dense, disordered skyrmion lattice \cite{Soumyanarayanan2017} and track its dependence on $T$ and $\kappa$ (see section 1 of the supplementary information for the evolution of magnetic textures across $T$ and $\kappa$).  
We explore the impact of this transition on the $THE$ by correlating the evolution of $\rho^{THE}$ across the $T-\kappa$ phase space. Fig.~2b displays $\rho^{THE}(T,\kappa)$ curves for individual multilayers studied as a function of $T$, while Fig.~2c shows $\rho^{THE}(\kappa)$ curves at a fixed $T$ with varying Fe/Co. The location of the maximum in $\rho^{THE}$ (see Fig. 2b,c) tracks the phase boundary between isolated skyrmions and a disordered skyrmion lattice tuned by $T$, $\kappa$. This confirms that the degree of $THE$ enhancement is closely linked to proximity to the phase boundary.
%%%%%%%%%%%%%
%%%%%%%%%%%%%%%

Here we note that the sign of $THE$ remains unchanged across the $T-\kappa$ phase diagram. However, $R_0$ changes from positive to negative with $T$ due to multiband transport \cite{raju2019evolution}. This crossover consistently follows the local maximum in $\rho^{THE}$ and the transition from isolated skyrmions to the lattice phase (for details see sections $4-6$ of the supplementary information). Such correlation between the local enhancement of $\rho^{THE}$, skyrmion configuration and sign reversal of $R_0$ suggests that these factors are influenced by systematic changes in the occupancy of the electronic bands while varying $T$ and the Fe/Co composition. To understand the mechanism responsible for the $R_0$ sign reversal, we employed a tight-binding model together with $ab-initio$ calculations of the electronic band structure of our multilayer stack. When electron like spin-up states and hole-like spin-down states are both present near the Fermi energy, a $T$ induced change in the individual carrier densities can lead to a sign change of $R_0$ due to compensation of the normal Hall signal from carriers with opposite charge. However, due to their opposing spin alignment, the interaction of these carriers with $B_{eff}$ from the skyrmions gives rise to a $THE$ which maintains the same sign from $T=5-300$~K. A complete quantitative analysis of the band structure is beyond the scope of this work; however, these experimental observations provide valuable insight into the links between thermodynamic stability and charge transport in chiral magnetic textures.

We now return to the enhancement of the $THE$. Within the critical region surrounding a second order phase transition, fluctuations of the  incipient order parameter ($\eta$) dominate the material response.  Hence, chiral spin fluctuations at all length scales may develop at the skyrmion lattice phase boundary~\cite{Nagaosa2013}. The spin chirality contribution to the $THE$ originates from a non-zero scalar triple product $[S{_i}\cdot(S{_j}\times S{_k})]$, where $S{_i}$, $S{_j}$, and $S{_k}$ form a cluster consisting of three fluctuating spins. Such clusters can generate Hall signals via an unconventional skew-scattering mechanism which can greatly exceed the contribution from a stable skyrmion configuration~\cite{Ishizuka2018,kato2019colossal,ishizuka2019theory,fujishiro2021giant}. The size of these clusters can be as small as several atomic spacings, which is much smaller than the skyrmion radius determined by the macroscopic magnetic interaction parameters in our films. Upon approaching the phase boundary, such fluctuation-induced clusters are expected to proliferate throughout the material, hence generating considerable topological charge~\cite{Ishizuka2018,kato2019colossal,ishizuka2019theory}. The correlation of these fluctuating spin clusters is associated with the formation/destruction of long-range order, i.e. a stable skyrmion lattice configuration.  Within this scenario, the spin correlation length of the skyrmion lattice (which for an infinite thermodynamic system would diverge at the transition) is distinct from the short length scale spin fluctuations responsible for a non-zero scalar spin chirality. A detailed microscopic understanding of this fluctuating spin chirality in relation to the phase transitions in chiral magnetic systems remain to be established theoretically \cite{fujishiro2021giant}. As we approach the transition the influence of such fluctuations should be revealed by power-law behavior in material properties, including the spin susceptibility and hence $THE$ \cite{wang2019spin}. The maximum observable $THE$ magnitude is expected to saturate at a value corresponding to the maximal chiral spin cluster density (imposed by the atomic spacing). Consequently, the power-law enhancement is truncated close to the transition and there is no $THE$ singularity.  
%%%%%%%%%%
%%%%%%%%%%%%%%%%

Our experimental observations reveal that the phase transition between isolated skyrmions and a disordered skyrmion lattice is sensitive to $\kappa$, $T$ and $H$. 
In the following we consider an effective temperature $T^{'}$ which accounts for the role of varying $T$ and $\kappa$ in controlling this phase transition. Figure 2d illustrates how the critical temperature $T_c$ and stability $\kappa_c$ defining the transition vary with $\kappa$ and $T$ respectively, in accordance with experiments (Fig.2a). The phase boundary between isolated skyrmions and a skyrmion lattice may be considered a smooth function $T_c(\kappa)$ \cite{leonov2018crossover,leonov2016properties,siemens2016minimal}. In the vicinity of a phase transition ($\kappa=\kappa_c\sim 1$, within the transition region, Fig.~2a) the boundary can be described as $T_{c}\cdot\kappa\approx$ constant and hence,

\begin{align*}%\right
    \frac{d\left(T_c\cdot\kappa\right)}{d\kappa}=\frac{d\left(constant\right)}{d\kappa} \Rightarrow \kappa\frac{dT_c}{d\kappa}+T_c=0 \Rightarrow \frac{dT_c}{d\kappa}\approx -T_{c},
\end{align*}
which in turn suggests an effective temperature $T'=T\cdot \kappa$ with a critical value $T_c' =T_c\cdot \kappa_c$ defining the transition. One may also consider the phase boundary in terms of varying $H$, however, the interdependence of $\kappa$ and $H$, and its variation with $T$ remain unclear. We therefore adopt the simpler picture presented in Fig.2d which allows a straightforward interpretation of our experimental observations in Fig.1 and 2a-c. 
 
%%%%%%%%%%%%%
%%%%%%%%%%%%%%%%%%%

Figure 3a shows a power-law behavior in $\Delta_{\rho/M}$ surrounding a critical value of $T^{'}_{c}\approx110\pm15$~K. This is consistent with the presence of a second order phase transition driven by $T$ and $\kappa$. According to Landau theory, the amplitude of $\eta$ grows as a power-law on the low-symmetry side of any second order transition (which in our case corresponds to $T'>T_c'$):

\begin{equation*}
    \eta \propto |T'-T_c'|^\beta,
\end{equation*}
whereas fluctuations of $\eta$ on both sides of the transition scale as:
\begin{equation*}
\langle (\Delta \eta)^2 \rangle \propto \frac{1}{|T'-T_c'|^\gamma}.
\end{equation*}
Extracting the critical exponents $\beta,\gamma$ from our data requires the identification of $\eta$ for the skyrmion lattice configuration. In the ordered phase of a chiral magnet, a helical wavevector which describes the spin rotation period emerges, capturing the symmetry broken by $\eta$ at the transition. Discrete Fourier transforms of a skyrmion lattice yield a strong peak in the structure factor at $k = l^{-1}$, where $l \propto 1/\sqrt{n_{sk}(MFM)}$ is the skyrmion lattice parameter and $k$ rises continuously from zero with the emergence of a lattice.  We therefore expect $\eta$ to scale with $\sqrt{n_{sk}(MFM})$, allowing analysis of the transition using the conventional scaling approach
\begin{equation}
{n_{sk}(MFM)}\approx {|T'-T'_c|^{2\beta}}, \qquad T'>T_c'.
\end{equation}

%%%%%%%%%%%%
%%%%%%%%%%%%%%%%%%%%

Fluctuations of $\eta$ in the vicinity of the phase transition create a topological charge and hence contribute an effective skyrmion density $n_{fl} \approx \langle (\Delta \eta)^2 \rangle$ ($n_{fl}\gg n_{sk}(MFM)$) to the total measured $n_{sk}(THE)$. Consequently, 

\begin{equation*}
    \Delta_{\rho/M}=\frac{n_{sk}(THE)}{n_{sk}(MFM)}=\frac{n_{sk}(MFM)+n_{fl}}{n_{sk}(MFM)}
    \approx \frac{n_{fl}}{n_{sk}(MFM)}, 
\end{equation*}

therefore we may model the power-law rise in $\Delta_{\rho/M}$ as follows:
\begin{align}
 & \Delta_{\rho/M} \sim \frac{1}{|T'-T'_c|^{\gamma}}, \qquad T'<T_c'
\\
 & \Delta_{\rho/M} \sim \frac{1}{|T'-T'_c|^{2\beta + \gamma}}, \qquad T'>T_c'
\end{align}

where the extra $2\beta$ in the exponent above $T'_c$ originates from the appearance of a stable skyrmion lattice as indicated in the eqn.$3$.

Fits to eqns.~$3$, $4$ to extract $\gamma$ and $\beta$ are shown as red lines in Fig.~3a (for $T^{'}<T_{c}^{'}$) and 3b respectively, where we use two independent data sets to obtain exponents $\gamma \approx 1.88\pm 0.43$, and $\beta \approx 0.31\pm0.05$. These values suggest a three dimensional Heisenberg spin system in which the exponents are modified by competing spin interactions in a quasi-two-dimensional environment.  This is in agreement with earlier studies of thin film magnets~\cite{LiPRL1992}, which consistently reveal an increased $\gamma$ and a reduced $\beta$ with respect to the three dimensional Heisenberg values $1.39$ and $0.36$ respectively~\cite{HolmPRB1993}. To independently cross-check our estimate of the exponents $\gamma$ and $\beta$ we also fit the discrepancy in the skyrmion lattice regime using eqn.~$5$ (Fig.~3a for $T^{'}>T_{c}^{'}$). We obtain the combination $(\gamma+2\beta)=2.61\pm0.44$, which is consistent with the $\gamma$ and $\beta$ extracted individually from our fluctuation analysis, adding credence to the validity of these critical exponents.

As discussed above, the finite size of the short lengthscale fluctuating spin clusters imposes an upper limit on the magnitude of $n_{fl}$, which is expected to saturate as the system approaches $T^{'}_{c}$.
Experimentally, we can only modulate $T^{'}$ in discrete steps, so we cannot tune $T^{'}$ continuously through $T^{'}_{c}$ to reveal the expected saturation in $n_{fl}$ and hence $THE$. However, our experimental $\Delta_{\rho/M}$ clearly demonstrates the anticipated power-law scaling on either side of $T^{'}_{c}$, systematically varying from $1-10^3$. Using eqn.$1$ and the values of $|R_{0}|\approx0.5-16$~n$\Omega$.cm/T measured in our multilayer films, we estimate that $\Delta_{\rho/M}$ should saturate at a value $~10^4 - 10^5$ as $|T^{'}-T^{'}_{c}|\rightarrow0$.  From the power-law trends shown in Fig. 3a, we deduce that this saturation is only visible for $|T^{'}-T^{'}_{c}|<20$~K, which lies beyond the minimum $|T^{'}-T^{'}_{c}|\approx40 \pm15$~K accessed during our experiments. It may therefore be possible to engineer a further THE increase of at least another order of magnitude, by tuning skyrmion-hosting multilayers more closely towards their isolated skyrmion/disordered lattice transition. Finally, we note that the presence of a finite population of skyrmions for $T^{'}<T^{'}_{c}$ is consistent with the nucleation-type transition by which the skyrmion lattice is established \cite{leonov2018crossover,leonov2016properties,siemens2016minimal}. Some of these isolated skyrmions may also be stabilised by local variations in the magnetic interactions due to disorder in our films.

%%%%%%%%%%%%%
%%%%%%%%%%%%%%%%
\section{Discussion}
Our results are summarised in Fig.~3c, which highlights the transition between isolated skyrmions and a disordered skyrmion lattice, determined by real-space imaging of the spin textures as well as magnetotransport. The critical parameter governing the transition $T_c(\kappa)$ is identified by three separate methods: the maximum in $\rho^{THE}$, $MFM$ imaging and the fluctuation-induced rise in $\Delta_{\rho/M}$ at $T_c^{'}=110\pm15$~K. All three data-sets display considerable overlap, indicating an active role of critical spin fluctuations in determining the magnitude of $THE$ in chiral magnetic films. The ensuing discrepancy of up to three orders of magnitude between $n_{sk}(THE)$ and $n_{sk}(MFM)$ in the vicinity of the phase transition may account for the widely-varying magnitudes of $\rho^{THE}$ values previously reported in technologically-relevant chiral magnetic films~\cite{zeissler2018discrete,raju2019evolution,wang2018ferroelectrically,tsymbal2018whirling,vistoli2019giant,he2018evolution,shao2019topological}.
Our material platform allows the $THE$ to be tuned from $n_{sk}(THE)\approx n_{sk}(MFM)$, which is typical for skyrmion crystals of B$20$ compounds \cite{Neubauer2009,Nagaosa2013,Spencer2018} to $n_{sk}(THE)> n_{sk}(MFM)$ in dilute skyrmion configurations characteristic of interfacial systems \cite{zeissler2018discrete,raju2019evolution,wang2018ferroelectrically,tsymbal2018whirling,vistoli2019giant,he2018evolution,shao2019topological}. This acute sensitivity of $\rho^{THE}$ to the magnetic skyrmion configuration indicates the crucial role of chiral spin fluctuations and opens a promising avenue towards controllable topological spintronics.

%%%%%%%%%%%%%%%%%%%%%%%%%%%%%%%%%%%%%%%

\bibliographystyle{naturemag}
%\bibliography{References}

%%%%%%%%%%%%%%%%%%%%%%%%%%%%
%%%%%%%%%%%%%%%%%%%%
\newpage
\FloatBarrier
\textbf{Acknowledgments:} This project in Singapore was supported by National Research Foundation (NRF) Singapore, under NRF Investigatorship programme (Ref. No. NRF-NRFI2015-04), the Ministry of Education (MOE) Singapore, Academic Research Fund (AcRF) Tier 2 (Ref. No. MOE2014-T2-1-050) and MOE AcRF Tier 3 (Ref. No. MOE2018-T3-1-002). M.R. thanks Data Storage Institute Singapore for sample growth facilities.  Analytical theory work at Ioffe Institute was supported by the Russian Science Foundation (Ref. No. 17-12-01265). K.S.D. and I.V.R. also thank the Foundation for the Advancement of Theoretical Physics and Mathematics “BASIS”. B.G., E.S. and I.M. acknowledge support by CRC/TRR 227 of Deutsche Forschungsgemeinschaft (DFG).

\textbf{Author contributions:} M.R. and C.P. conceived the research and coordinated the project. M.R. carried out the deposition of films, magnetotransport measurements and data analysis. A.P.P. provided input to the analysis of the transport data. A.Y. performed low temperature MFM experiments and analysed the data with O.M.A. N.K.D. performed room temperature MFM experiments and analysed the data. K.S.D. and I.V.R. provided theoretical insights into the topological Hall effect in disordered skyrmion systems. B.G. and I.M. provided theoretical insights into the Hall effects in multiband systems for which E.S. provided the ab-initio calculations. M.R. and C.P. wrote the manuscript with discussions and contributions from all authors. 

\textbf{Additional Information:} Supplementary Information accompanies this paper.

\textbf{Competing Interests:} The authors declare no competing interests.

\textbf{Data availability:} The authors declare that the data supporting the findings of this study are available within the paper, and its supplementary information. 

\textbf{Correspondence:} Correspondence and requests for materials should be addressed to:\\ mraju@ntu.edu.sg/christos@ntu.edu.sg

\newpage
\FloatBarrier
\textbf{Methods}\\
\textbf{Film deposition:} Thin film multilayers consisting of Ta($30$)/Pt($100$)/[Ir($10$)/Fe($x$)/Co($y$)/Pt($10$)]{$_N$}/Pt($20$) (numbers in the parentheses are layer thickness in \AA~and $N$ refers to the number of times the Ir/Fe/Co/Pt stack is repeated, $x, y$ are varied between $0-6$ \AA~and $5-10$ \AA~respectively) were deposited on thermally oxidised Si wafers by dc magnetron sputtering at room temperature, using a Chiron ultra-high vacuum multi-source system. The optimal growth rates for individual layers are Ta: $0.55$ \AA/s, Pt: $0.47$ \AA/s, Ir: $0.12$ \AA/s, Fe: $0.13$ \AA/s, Co: $0.2$ \AA/s. The base vacuum of the sputter chamber is $1\times10^{-8}$~Torr and an argon gas pressure of $1.5\times10^{-3}$~Torr is maintained during sputtering.     

\textbf{Electrical transport:} The magnetotransport measurements were performed using a custom-built variable temperature insert (VTI) housed in a high-field magnet, complemented by a Quantum Design Physical Property Measurement System (PPMS). Current densities as low as $10^4$~A/m$^2$ at $33$~Hz were used to avoid current driven perturbation of spin textures. Detailed analysis for the extraction of the topological Hall resistivity ($\rho^{THE}(H)$) can be found in our earlier works \cite{raju2019evolution,Soumyanarayanan2017}. The non-zero value of $\rho^{THE}$ at $H>H_S$ serves as a conservative estimate of the error bar in the extracted $\rho^{THE}(H)$ which is $\leq2$~n$\Omega$.cm. This includes the systematic errors involved in data analysis.

\textbf{Magnetization:} Magnetization measurements were performed in the region $T=5-300$~K and a magnetic field of $\pm4$~T using superconducting quantum interference device (SQUID) magnetometry, in a Quantum Design Magnetic Property Measurement System (MPMS) (see section 2 of the supplementary informaton). $M(H)$ loops were recorded with the applied field in-plane (hard axis) and out of the film plane (easy axis). The saturation magnetization $M_S$ and the difference in saturation fields $H_S$ along the easy and hard axes were used to estimate the effective magnetic anisotropy $K_{eff}$ \cite{Soumyanarayanan2017,raju2019evolution}.     

\textbf{Magnetic interactions:} $M_S$ and $K_{eff}$ were determined from SQUID magnetometry measurements. The detailed methods for estimation of exchange stiffness $A$ and $DMI$ were  reported in our previous work \cite{Soumyanarayanan2017}. In this work, we use scaling laws, $\frac{A(T)}{A(T=5~K)}=\left[ \frac{M_S{(T)}}{M_S{(T=5~K)}}\right]^{1.5}$ and $\frac{D(T)}{D(T=5~K)}=\left[ \frac{M_S{(T)}}{M_S{(T=5~K)}}\right]^{1.5}$ involving the $T$-dependent saturation magnetization $M_S(T)$ to estimate $A(T)$, $D(T)$, and the resulting $\kappa(T)$. Details can be found in section 2 of the supplementary information.

\textbf{Magnetic force microscopy ($MFM$):} Room temperature $MFM$ experiments were carried out using a Veeco Dimension $3100$ Scanning Probe Microscope. The $MFM$ tips used (Nanosensors SSS-MFMR ) were $\approx30$~nm in diameter, with low coercivity ($\approx12$~mT) and ultralow magnetic moment ($\approx80$~emu/cc). Samples were initially saturated in out of plane ($OP$) direction using fields up to $H=-0.5$~T. The measurements were performed in $OP$ fields starting from $H=0$ and incrementally approaching $H=+H_S$, with a typical tip height of $20$~nm. The field evolution of spin textures is presented in section 1 of the supplementary information. 
Low $T$ ($5-200$~K) $MFM$ imaging is carried out using a cryogenic frequency modulated $MFM$ system \cite{raju2019evolution}. We used two commercial probes by Team Nanotec, model ML3 ($35-40$~nm Co alloy coating), with $f_0\approx 75$~kHz and $k_0\approx 1$~N/m. The sample was first stabilized at a given temperature and then magnetized in the $OP$ direction by applying $H>H_S$. After saturation, $MFM$ images were acquired at various field values as $H$ was swept from $-H_S$ to $+H_S$. The details of the $MFM$ image analysis are reported in our earlier reports \cite{Soumyanarayanan2017, raju2019evolution,Yagil2018}.

\textbf{Calibration of applied magnetic field:} Extensive field calibrations were performed using a reference Hall sensor and a reference Palladium sample to minimize the field offsets between different experimental setups. Sensor details, calibration data together with analysis for the extraction of $\rho^{THE}(H)$ are reported in our earlier works \cite{Soumyanarayanan2017,raju2019evolution}. An additional field-calibration crosscheck is performed for our $MFM$ images  by estimating the magnetization from the image itself. Here the magnetization of the image is estimated as the ratio of the effective area of the imaged surface polarized along the applied field to the total area of the image. The magnetization of the $MFM$ image is then located on $M(H)$ data recorded by SQUID magnetometer.   

\newpage
\FloatBarrier
\begin{figure}
    \centering
    \includegraphics[width = 0.9\textwidth]{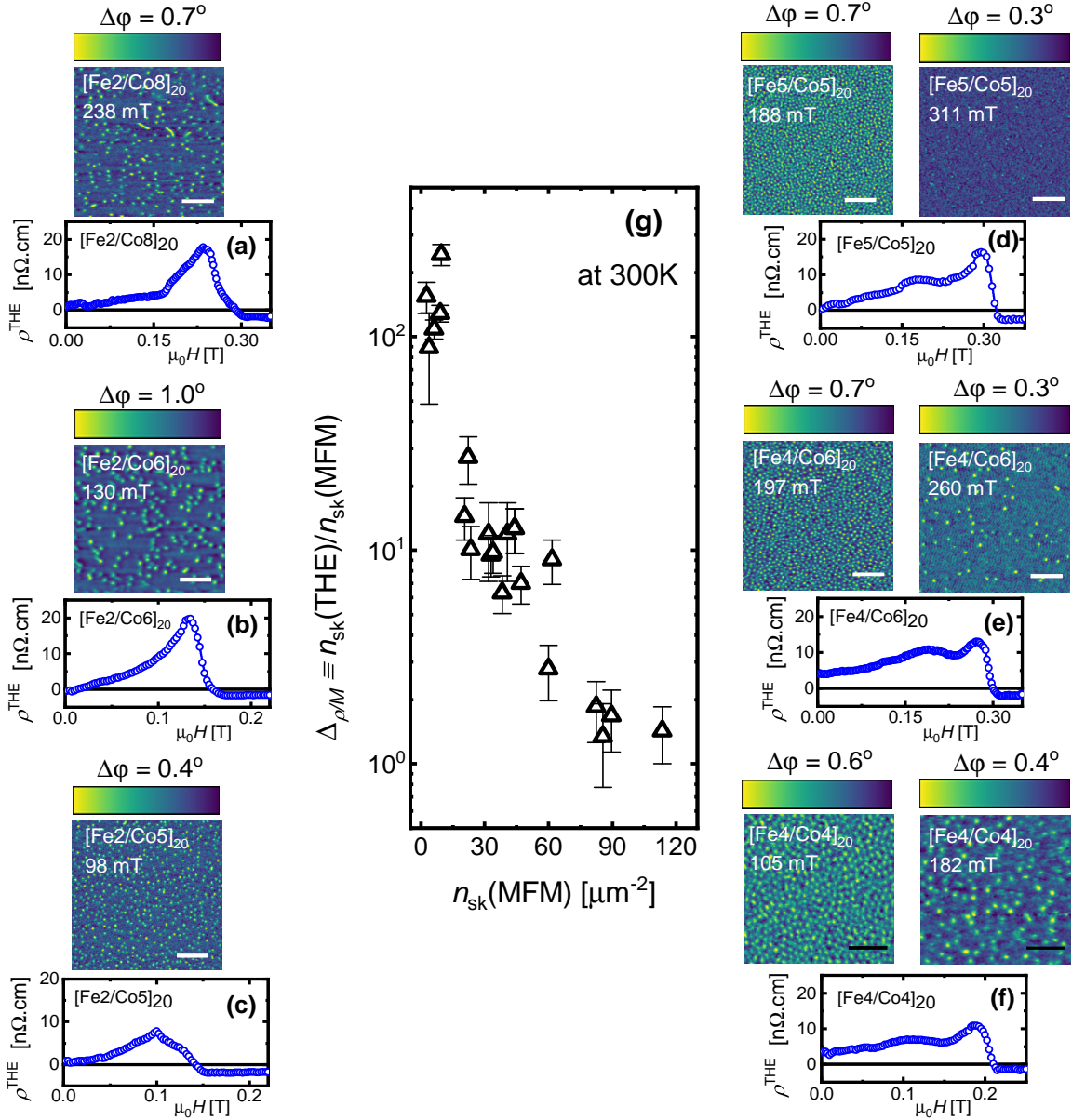}
   \caption{
   {\textbf{Enhanced topological Hall resistivity ($\rho^{THE}$) for isolated skyrmions compared with dense skyrmion arrays:} \textbf{(a)-(c)} $\rho^{THE}(H)$ of isolated skyrmion configurations observed in [Fe2/Co8]$_{20}$, [Fe2/Co6]$_{20}$ and [Fe2/Co5]$_{20}$. \textbf{(d)-(f)} $\rho^{THE}(H)$ of dense skyrmion arrays observed in [Fe5/Co5]$_{20}$, [Fe4/Co6]$_{20}$, and [Fe4/Co4]$_{20}$. \textbf{(g)} $\Delta_{\rho/M}\equiv~n_{sk}(THE)/n_{sk}(MFM)$ estimated from transport and imaging experiments shown in (a-f) and in section 1 of the supplementary information. $n_{sk}(THE)$ is given by $\rho^{THE}/(P\cdot R_0\cdot \Phi_0)$ with $P=0.56$ and $n_{sk}(MFM)$ is estimated from $MFM$ images; details of our image analysis methods can be found in our earlier reports \cite{Soumyanarayanan2017,raju2019evolution,Yagil2018}. The detailed evolution of magnetic textures with $H$ can be found in section $1$ of the supplementary information. Scale bar in white ($MFM$ images corresponding to panels (a)-(e)) are $1$~$\mu$m  and in black ($MFM$ images corresponding to panel (f)) $0.5$~$\mu$m. All the images are acquired at a scan height of $20$~nm. The color scale represents the phase shift in the $MFM$ signal due to the magnetic force acting on the tip. The $\rho^{THE}(H)$ profiles show a small non-zero offset $\leq2$~n$\Omega\cdot$cm above the saturation field of magnetization ($H>H_s$): this is a systematic offset due to our fitting procedure (which is designed to avoid any ``false positive'' $THE$ detection) and hence reflects the maximum error in the magnitude of $\rho^{THE}(H)$ shown in Fig.$1$g. }}
    \label{figure 1}
\end{figure}

\newpage
\FloatBarrier
\begin{figure}
    \centering
    \includegraphics[width = 1.0\textwidth]{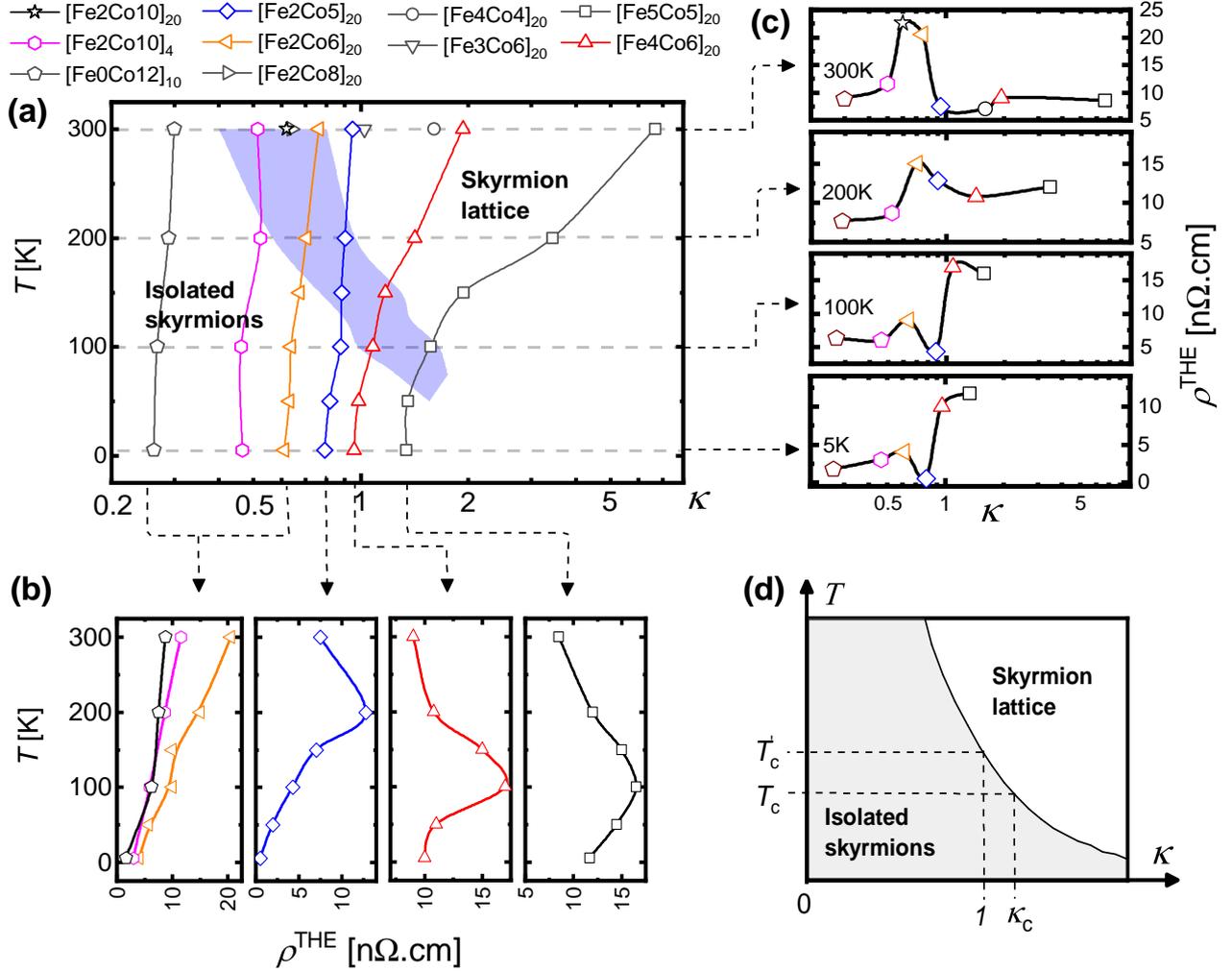}
    \caption{{\textbf{Evolution of $\rho^{THE}$ across the phase transition between isolated skyrmions and a disordered skyrmion lattice in $T-\kappa$ space}. \textbf{(a)} Phase diagram of skyrmion configurations in $T-\kappa$ space obtained by varying $T$ and Fe/Co composition. The shaded area in blue is the transition region from isolated skyrmions to a disordered skyrmion lattice revealed by $MFM$ imaging. The spatial evolution of skyrmion configuration is analysed using Delaunay triangulation: details of the analysis including the nearest-neighbour (NN) coordination number and NN angular orientation can be found in section $1$ of the supplementary information \cite{Soumyanarayanan2017}. Details of our estimation of the $T$-dependent $\kappa$ are presented in section $2$ of the supplementary information. \textbf{(b)} Evolution of $\rho^{THE}$ with $(T, \kappa)$. Arrows with dotted lines point to the specific samples in panel (a) for which $\rho^{THE}$ is studied as a function of $T$. \textbf{(c)} Evolution of $\rho^{THE}$ with $\kappa$ at a fixed $T$. Arrows with dotted lines point to the various samples in panel (a) for which $\rho^{THE}(\kappa)$ is studied at a fixed $T$. \textbf{(d)} Schematic $T-\kappa$ phase diagram illustrating the phase boundary $T_{c}(\kappa)$ and the use of the effective temperature $T^{'}=T\cdot\kappa$ to describe the phase transition. The error bars on $\rho^{THE}$ in panels (b) and (c) are $\leq2$~n$\Omega\cdot$cm.}
    }
    \label{fig:2}
    
\end{figure}

\newpage
\FloatBarrier
\begin{figure}
    \centering
    \includegraphics[width = 1.0\textwidth]{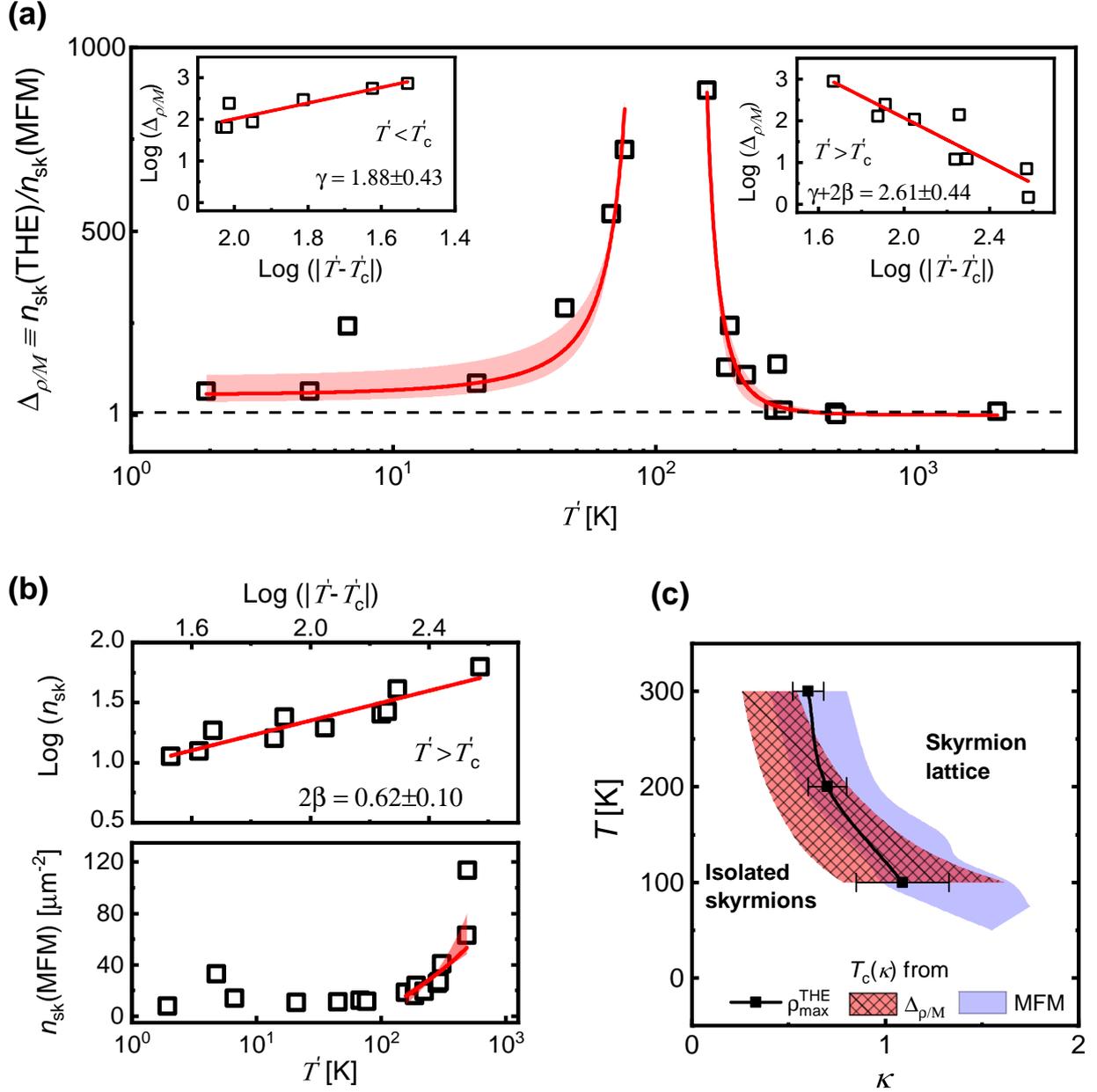}
    \caption{\textbf{Insights into the critical behavior captured by the topological Hall effect ($THE$)}. \textbf{(a)} Power-law behavior of $\Delta_{\rho/M}\equiv~n_{sk}(THE)/n_{sk}(MFM)$ with effective temperature $T^{'}=T\cdot\kappa$, and a critical point at $T_c^{'}=110\pm15~K$. Insets show the estimation of critical exponents, $\gamma$ for $T^{'}<T^{'}_{c}$ and $\gamma +2\beta$ for $T^{'}>T^{'}_{c}$. \textbf{(b)} Evolution of $n_{sk}(MFM)$ with $T^{'}$, showing a power-law behavior for $T^{'}>T^{'}_{c}$ and the estimation of the critical exponent $2\beta$. \textbf{(c)} Identification of the critical region $T_{c}(\kappa)$ using three different approaches, namely, the local maximum in $\rho^{THE}$ with $T$ and $\kappa$, the power-law rise in $\Delta_{\rho/M}$ and the transition region (from isolated skyrmions to a disordered skyrmion lattice) indicated by $MFM$. The error bar in the estimated exponents reflects the error in the slope of the linear fits shown in the insets to panel $3$a and $3$b. The same error bar is reflected through the shaded regions in the power-law fits. Error bars on $\rho^{THE}_{max}$ in panel $3$c reflect the variation in the peak position of $\rho^{THE}$ between Fig.$2$b and $2$c. The shaded region for $T_{c}(\kappa)$ from $\Delta_{\rho/M}$ in panel $3$c reflects a conservative estimate of $\pm35\%$ variation in $\kappa$ resulting from an upper bound in the estimation of the exchange ($A$) and $DMI$~(D) constants as well as $K_{eff}$ from magnetization measurements \cite{Soumyanarayanan2017}.}
    
    \label{fig:3}
\end{figure}

\end{document}